\documentclass[aps,preprint]{revtex4}%
\usepackage{amsfonts}
\usepackage{amsmath}
\usepackage{amssymb}
\usepackage{graphicx}%
\providecommand{\U}[1]{\protect\rule{.1in}{.1in}}

\begin{document}
\title{Approximate Equations of Motion for Compact Spinning Bodies in General Relativity}
\author{James L. Anderson}
\affiliation{Stevens Institute of Technology}
\affiliation{Hoboken, NJ 07030}
\pacs{PACS number}
\date{November 14, 2005}

\begin{abstract}
Approximate equations are derived for the motion of a gyroscope on the earth's
gravitational field using the Einstein, Infeld, Hoffmann surface integral
method. This method does not require a knowledge the energy-momentum-stress
tensor associated with the gyroscope and uses only its exterior field for its
characterization. The resulting equations of motion differ from those of
previous derivations.

\end{abstract}
\maketitle

\bigskip

Already by 1927 Einstein had realized that the field equations of general
relativity contained information about the motion of the sources of the
gravitational field.\cite{EG} A complete working out of this possibility was
achieved by Einstein, Infeld and Hoffmann\cite{EIH} (EIH) and elaborated later
in two papers by Einstein and Infeld\cite{EI}. Their method used only the
source-free field equations of general relativity and avoided any
consideration of the interiors of the sources, assuming only that they are
compact, \textit{i.e. }their sizes are small compared to the distance between
them and that their velocities are small compared to the speed of light $c$.
The nature of the sources is characterized by their exterior fields alone. The
main drawback to the EIH procedure is that it requires a large amount of
tedious calculation.

An alternate approach to deriving equations of motion was introduced by
Fock\cite{Fock} and later developed further by Papapetrou\cite{Pap}. This
second method takes into account the interiors of the sources and makes use of
the conservation laws
\begin{equation}
T^{\mu\nu}{}_{;\nu} \label{c1}%
\end{equation}
where $T^{\mu\nu}$ is the matter tensor associated with the source. (Here and
in what follows Greek indices take the values $0,1,2,3,$ Latin indices take
the values 1,2,3, the Einstein summation convention is assumed, a comma
denotes ordinary differentiation and a semicolon denotes covariant
differentiation.) While this method requires somewhat less calculations, it
has the drawback that one must specify $T^{\mu\nu}$. \ Papapetrou\cite{Pap2}
later used the Fock approach to derive equations of motion for `test' bodies
moving in an external gravitational field and applied it to the case of a
spinning test body. However, the equations obtained by him were insufficient
to determine completely the dynamical variables $X^{\mu}$ and $S^{\mu\nu}$,
the position and spin angular momentum of the body, appearing in them. As a
consequence it was necessary to impose restrictions from the outside on the
components of $S^{\mu\nu}$ in order to close the system of equations. These
restrictions all require the vanishing of its space-time components (the i0
components) but differ in which frame of reference these components vanish.
Various authors\cite{VO} have proposed different restrictions leading to
different equations of motion for $S^{\mu\nu}.$ In a completely different
approach using the Principle of Equivalence, Weinberg\cite{Wein} has obtained
yet another set of equations for this variable that differs from those based
on the Papapetrou equations. It should be pointed out that neither of these
derivations make full use of the Einstein field equations - Papapetrou used
them only to derive the conservation laws exhibited in equation (\ref{c1}) and
Weinberg did not use them at all. A resolution of these different results has
taken on a certain urgency with the launch of the NASA-Standard Gravity Probe
B (GPB) designed to measure the change in orientation of a small gyroscope in
earth orbit. This experiment was first suggested by Schiff\cite{Sch} in 1960,
\ whose calculations were based on the spin equations of motion obtained using
the supplementary conditions proposed by Corinaldesi and Papapetrou (CP).

In an attempt to resolve these issues I have undertaken to derive approximate
equations of motion for compact spinning bodies using methods similar to the
ones used by EIH. Today two developments have made it possible to perform the
calculations needed for this purpose without undo labor - high speed personal
computers and symbolic manipulation programs such as Mathematica\cite{Math}
and Maple\cite{Maple} together with the wonderful program grtensor\cite{grt}.
With these tools it is now possible to obtain equations of motion in a matter
of minutes with the assurance of correctness that would have required days or
even weeks to perform by hand.

Since the equations of motion obtained using the EIH approach are only
approximate, it is necessary to specify the system one wishes to apply these
equations to, in this case a compact gyroscope orbiting the earth, and to
identify the small dimensionless parameters associated with this system that
will be used in the expansions employed. The gyroscope used in GPB ( there are
actually four of them) consists of an almost perfect sphere of fused quartz
with a radius $r_{g}=.019$ m and a mass $m=.075$ kg. and an initial angular
velocity $\omega=27000$ rad/s. The gyroscope was launched into a near perfect
circular polar orbit (eccentricity = .0014) of radius $R=$ 7027 km. These
values will determine the relative importance of the terms in the approximate
expression for the gravitational field to be used to evaluate the surface
integral terms that arise in the EIH approach.

The form of the field equations for the gravitational field $g_{\mu\nu}$ to be
used here are due to Landau and Lifshitz\cite{LL}. Exterior to the field
sources they have the form%
\begin{equation}
U^{\mu\nu\rho}{}_{,\rho}=-g\,t_{LL}{}^{\mu\nu}\text{,} \label{c2}%
\end{equation}
where%
\begin{equation}
U^{\mu\nu\rho}=-U^{\mu\rho\nu}=({\small 1/16\pi})\{-g(g^{\mu\nu}g^{\rho\sigma
}-g^{\mu\rho}g^{\nu\sigma})\}_{,\sigma}\text{ ,} \label{c3}%
\end{equation}
$g=$ det$(g_{\mu\nu})$ and $t_{LL{}}{}^{\mu\nu}$ is the Landau-Lifshitz
pseudotensor. Because of the antisymmetry of $U^{\mu\nu\rho}$ in its last two
indices, it follows that $U^{\mu rs}{}_{,s}$ is a three-dimensional curl and
therefore when equation (\ref{c2}) is integrated over a two-surface in a $t=$
constant hypersurface, one gets%
\begin{equation}%
{\textstyle\oint\nolimits_{S}}
(U^{\mu r0}{}_{,0}+gt_{LL{}}{}^{\mu r})n_{r}dS\text{ ,} \label{c4}%
\end{equation}
where $n_{r}$ is a unit surface normal. In a like manner one gets%
\[%
{\textstyle\oint\nolimits_{S}}
\{(x^{\mu}U^{\nu r0})_{,0}-(x^{\nu}U^{\mu r0})_{,0}+gx^{\mu}t_{LL}{}^{\nu
r}-gx^{\nu}t_{LL}{}^{\mu r}\text{ \ \ \ \ \ \ \ \ \ \ \ \ \ }%
\]%
\begin{equation}
\text{ \ \ \ \ \ \ \ \ \ \ \ \ \ \ \ \ \ \ \ \ \ \ \ \ \ \ \ \ \ \ \ \ \ \ }%
+({\small 1/16\pi})\{g(g^{\nu r}g^{\mu0}-g^{\mu r}g^{\nu0})_{,0}%
\}n_{r}dS=0\text{.} \label{c5}%
\end{equation}
It is these two last equations that are used in the EIH procedure to obtain
equations of motion.

In order to use equations (\ref{c4}) and (\ref{c5}) it is necessary to obtain
solutions of the field equations corresponding to the type of system being
considered. Since in the case of GPB no such exact solution exists it is
necessary to use approximate ones. In the case of GPB the system consists of
two bodies, the earth and the gyroscope. Since the ratio of the masses $M$ and
$m$ of the earth and the gyroscope is 1.25$\times10^{-26}$ it is clear that we
can ignore completely any effect the gyroscope has on the earth's motion. We
can therefore take the earth to be at rest at the origin of an inertial frame
characterized by coordinates \{$ct,x,y,z$\}. At the location of the gyroscope
the earth's field has the dimensionless value $MG/Rc^{2}=$ $6.5\times10^{-10}$
where $G$ is the Newtonian gravitational constant and M is the mass of the
earth. The \ gyroscope's contribution to the gravitational field consists of
three parts: its static part $mG/r_{g}c^{2}=2.9\times10^{-27}$, an induction
part $mGV/r_{g}c^{3}=7.3\times10^{-32}$, where $V=7.5\times10^{3}$ m/s is the
orbital velocity of the gyroscope, and a spin contribution. The latter
contribution depends on the gyroscope's spin angular momentum $S=0.29$ kg
m$^{2}$/s and is given by $SG/r_{g}^{2}c^{3}=2.1\times10^{-33}$. From these
numbers we can form the three small dimensionless parameters to be used in the
construction of the approximate gravitational field. The first of these is the
slowness parameter $\epsilon=V/c=2.5\times10^{-5}$. \ The second one is the
ratio of the gyroscope's monopole field to that of the earth at the surface of
the gyroscope, $\epsilon{\small 1=mR/Mr}_{g}=4.6\times10^{-18}$. Finally, the
third small parameter is the ratio of the spin angular momentum of the
gyroscope to its orbital angular momentum $\epsilon{\small 2=}=S/mVR=7.4$
$\times10^{-11}$ .

Taking these considerations into account, using units with $G=c=1$ and
measuring masses in units of $M$ and lengths in units of $R$, the components
of the gravitational field $g_{\mu\nu}$ have the form%
\begin{subequations}
\begin{equation}
g_{00}=1-2\epsilon^{2}\frac{M}{r}-2\epsilon^{2}\epsilon{\small 1}\frac
{m}{r{\small 1}}-2\epsilon^{4}\epsilon{\small 2\gamma}_{i}V_{i} \label{6a}%
\end{equation}

\begin{equation}
g_{ij}=\delta_{ij}(-1-2\epsilon^{2}\frac{M}{r}-2\epsilon^{2}\epsilon
{\small 1}\frac{m}{r{\small 1}})-\epsilon^{4}\epsilon{\small 2(\gamma}%
_{i}V_{j}+\gamma_{j}V_{i}) \label{6b}%
\end{equation}

and%
\begin{equation}
g_{i0}=2\epsilon^{3}\epsilon{\small 1}\frac{mV_{i}}{r{\small 1}}+\epsilon
^{3}\epsilon{\small 2}\,\gamma_{i} \label{38}%
\end{equation}
where%
\end{subequations}
\begin{equation}
\gamma_{i}=\varepsilon_{ijk}\frac{x{\small 1}_{j}s_{k}}{r{\small 1}^{3}}
\label{39}%
\end{equation}
and where $r^{2}=x^{2}+y^{2}+z^{2}$ is the distance from the earth's center
with coordinates $\{0,0,0\}$ to the field point $\{x,y,z\}$, $r{\small 1}%
^{2}=x{\small 1}_{1}^{2}+x{\small 1}_{2}^{2}+x{\small 1}_{3}^{2}$ is the
distance from the center of the gyroscope with coordinates$\{R_{1},R_{2}%
,R_{3}\}$ to this field point and $\varepsilon_{ijk}$ is the antisymetric
density with values +1 or -1 depending on whether $ijk$ is an even or odd
permutation of 123 and zero otherwise. Here only those terms that are needed
to determine the lowest order equations of motion for the $s_{k}$ have been
included in these expressions for the approximate components of $g_{\mu\nu}$.
It is to be noted that the gyroscope monopole and dipole contributions to
$g_{\mu\nu}$ are taken to have the same effective centers so that no
supplementary conditions are needed to determine the time dependent of the
$s_{k}$.

To obtain the above expressions for the gravitational field one makes use of
the field of a stationary spinning, spherically symmetric, body given by
$g_{i0}$ above. Since the GPB gyroscopes are moving in the earth's
gravitational field, it is necessary to boost the static field to the velocity
of the moving gyroscope. This boost is responsible for the terms in the
expressions for $g_{00}$ and $g_{ij}$ above that depend on $s_{i}.$ In
addition it is necessary that these fields satisfy the harmonic coordinate
conditions to an accuracy that insures that the $g_{\mu\nu}$ are in fact
approximate solutions of the Einstein field equations. In the present case
this requirement is satisfied if
\begin{equation}
(\sqrt{-g}g^{\mu\nu})_{,\nu}=\mathcal{O}(\epsilon^{4}\epsilon{\small 1)}\text{
} \label{40}%
\end{equation}
This will be the case provided that
\begin{equation}
s_{i,0}=\mathcal{O}(\epsilon^{3})\text{ } \label{41}%
\end{equation}
Finally,it is necessary to discuss the dependence on time of the dynamical
variables $R_{i}$ and $s_{i}$. In their original paper, EIH introduced what
they called the slow-motion approximation by assuming that the source
coordinates depended on the time $t$ through the combination $\epsilon t$.
Their procedure is equivalent to what is known today as a multiple time
formalism (\cite{JLA}) and will be used in what follows. This being the case,
condition (\ref{41}) can be satisfied if one assumes that%
\begin{equation}
s_{i}=s_{0i}+\epsilon s_{1i}(\epsilon t) \label{43}%
\end{equation}
where $s_{0i}$ is independent of $\epsilon t$ . (In higher orders of
approximation $s_{0i}$ will in general depend on $\epsilon^{3}t$ .)

All that remains is to substitute the above expressions for $g_{\mu\nu}$ into
the surface integrals in equations (\ref{c4}) and (\ref{c5}) and evaluate the
integrals over a sphere surrounding the gyroscope. Most of the terms so
obtained will depend on the radius of the sphere chosen and so must cancel as
a consequence of the field equations (\ref{c2}). Those terms that are
independent of the sphere radius must vanish as a consequence of the motion of
the sources, here the gyroscope, and hence are the desired equations of
motion. Evaluating the surface integrals in equation \ref{c4} to
$\mathcal{O(\epsilon}^{4}\epsilon{\small 1)}$ yields the Newtonian equations
of motion for a particle moving in the earth's gravitational field%
\begin{equation}
\mathbf{R}^{%
\acute{}%
\acute{}%
}=-\frac{M\mathbf{R}}{R^{3}} \label{53}%
\end{equation}
where a prime denotes differentiation with respect to $\epsilon t$.

To get equations for the spin variables $s_{i}$ it is necessary to evaluate
the surface integrals in equation(\ref{c5}) to $\mathcal{O(\epsilon}%
^{6}\epsilon{\small 2})$, a task that would have been beyond my abilities to
perform without the help of Mathematica and grtensor. One finds that these
equations can be written in vector form as%
\begin{equation}
d\mathbf{s}_{1}\mathbf{/}d(\epsilon t)=\frac{1}{10}\frac{M}{R^{3}%
}\{-(\mathbf{s}_{0}\mathbf{\cdot R}^{%
\acute{}%
}\mathbf{)R+19(s}_{0}\mathbf{\cdot R)R}^{%
\acute{}%
}+16\mathbf{s}_{0}(\mathbf{R\cdot R}^{%
\acute{}%
})\} \label{44}%
\end{equation}
These equation are to be compared to the ones obtained by Cornaldesi and
Papapetrou given by%
\begin{equation}
d\mathbf{s/}dt=2\frac{M}{R^{3}}\{-\frac{1}{2}(\mathbf{s\cdot}\overset{\cdot
}{\mathbf{R}}\mathbf{)R+(s\cdot R)}\overset{\cdot}{\mathbf{R}}+2\mathbf{s}%
(\mathbf{R\cdot}\overset{\cdot}{\mathbf{R}})-\frac{3}{2}\frac{(\mathbf{R\cdot
}\overset{\cdot}{\mathbf{R}})}{R^{2}}(\mathbf{s\cdot R)R}\} \label{45}%
\end{equation}
and Weinberg's equation%
\begin{equation}
d\mathbf{s/}d(t)=\frac{M}{R^{3}}\{-2(\mathbf{s\cdot}\overset{\cdot}%
{\mathbf{R}}\mathbf{)R+(s\cdot R)}\overset{\cdot}{\mathbf{R}}-2\mathbf{s}%
(\mathbf{R\cdot}\overset{\cdot}{\mathbf{R}})\}\text{ .} \label{46}%
\end{equation}
where a dot over a quantity denotes differentiation with respect to $t$.
\ These latter two sets of equations are considered to be exact by their
authors but can be solved approximately by substituting for $\mathbf{s}$ the
expression given for it in equation (\ref{43}). It is clear that they will
give different results than those obtained from equations (\ref{44}). In all
of these equations the dependence of $\mathbf{R}$ on $t$ is gotten by solving
the Newtonian equations of motion (\ref{53}).

In the case of a circular motion in the xy-plane we can take, in equation
(\ref{44}),
\begin{equation}
R_{1}=R\cos(\epsilon\omega t)\text{ and }R2=R\sin(\epsilon\omega t)\text{.}
\label{47}%
\end{equation}
where $\omega$ is the angular velocity of the gyroscope in its orbit. To find
the secular change in $\mathbf{s}$ with time we can average Equation.(\ref{44}%
) over an orbital period $T=2\pi/\omega$. If one takes $\mathbf{s}_{0}$ to lie
in the plane of the orbit and the xy-axes are chosen so that $s_{01}=s_{0}$
and $s_{02}=0$, the resultant change $\Delta s_{1}$ is given by
\begin{equation}
\Delta\,s_{11}=0 \label{48}%
\end{equation}
and%
\begin{equation}
\Delta\,s_{12}=2\pi s_{0}M/R\text{ } \label{49}%
\end{equation}
so that the angular change $\Delta\theta$ in the direction of $\mathbf{s}$ is
given by
\begin{equation}
\Delta\,\theta=2\pi M/R\text{ .} \label{50}%
\end{equation}
A similar analysis yields a value $\Delta\theta=M/2R$ for the CP equations and
$3M/2R$ for the Weinberg equations. Why the difference in the three
results?\ In the case of the Papapetrou-Corinaldesi equations the authors made
assumptions concerning the matter tensor $T^{\mu\nu}$ that are not justified
and Weinberg relied on the Principle of Equivalence and identified the space
part of a four-vector with an axial three- vector.

It is also possible to take account of the spin-spin interaction between the
gyroscope and the earth's rotation about its axis. To do so it is first
necessary to introduce a fourth small parameter $\epsilon{\small 3}%
=\Omega/\omega=6.8\times10^{-2}$, where $\Omega$ is the angular velocity of
the earth, into our expansions. The gravitational field of the earth's
rotation can be taken account of by adding to the expression (\ref{38}) for
$g_{i0}$ a term
\begin{equation}
g%
\acute{}%
_{i0}=\epsilon^{3}\epsilon{\small 3}\varepsilon_{ijk}\frac{R_{j}S_{k}}{R^{3}}
\label{52}%
\end{equation}
where $S_{k}=I\,\Omega_{k}$ is the earth's angular momentum and $I$ is its
moment of inertia about its axis of rotation. With this addition to the
gravitational field, the surface integral (\ref{c5}) introduces an additional
term $\mathbf{\tau}_{S}$ in the equation of motion (\ref{44}) for
$\mathbf{s}_{1}$ given by%
\begin{equation}
\mathbf{\tau}_{S}=\frac{\epsilon{\small 3}}{2R^{5}}\left\{  R^{2}%
(\mathbf{s}_{0}\times\mathbf{S)-3(S\cdot R)(s}_{0}\times\mathbf{R}\right\}  .
\label{54}%
\end{equation}

For a circumpolar orbit with xy-axes now chosen so that $\mathbf{S}%
=\{S,0,0\}$, assuming that $\mathbf{s}_{0}=\{s_{0}\cos(\varphi),s_{0}%
\sin(\varphi),0\}$ and after averaging over an orbital period one finds an
additional change in $\mathbf{s}_{1}$ given by%
\begin{equation}
\Delta\,s_{13}=\frac{2\pi}{4R^{3}}I\frac{\Omega}{\omega}s_{0}\sin
(\varphi)\label{55}%
\end{equation}
with a corresponding angular change in the direction of $\mathbf{s}$ given by%
\begin{equation}
\Delta\,\theta=\frac{2\pi}{4R^{3}}I\frac{\Omega}{\omega}\sin(\varphi
).\label{56}%
\end{equation}
It is amusing to think that if this additional change in direction could be
measured with enough accuracy one could use the result to determine $I$ by
assuming that general relativity was the correct theory of gravity.

\bigskip

* e-mail address: jlanders@stevens.edu

\end{document}